# ASYMMETRIC PEER INFLUENCE IN SMARTPHONE ADOPTION IN A LARGE MOBILE NETWORK

Qiwei Han, Department of Engineering and Public Policy, Carnegie Mellon University, 5000 Forbes Avenue, Pittsburgh, PA 15213, USA and Instituto Superior Técnico, University of Lisbon, Av. Rovisco Pais, Lisbon 1049-001, Portugal, qiweih@cmu.edu

Pedro Ferreira, Heinz College and Department of Engineering and Public Policy, Carnegie Mellon University, 5000 Forbes Avenue, Pittsburgh, PA 15213, USA, pedrof@cmu.edu

João Paulo Costeira, Instituto Superior Técnico, University of Lisbon, Av. Rovisco Pais, Lisbon 1049-001, Portugal, jpc@isr.ist.utl.pt

## Abstract

*Understanding adoption patterns of smartphones is of vital importance to telecommunication managers in today's highly dynamic mobile markets. In this paper, we leverage the network structure and specific position of each individual in the social network to account for and measure the potential heterogeneous role of peer influence in the adoption of the iPhone 3G. We introduce the idea of core-periphery as a meso-level organizational principle to study the social network, which complements the use of centrality measures derived from either global network properties (macro-level) or from each individual's local social neighbourhood (micro-level). Using millions of call detailed records from a mobile network operator in one country for a period of eleven months, we identify overlapping social communities as well as core and periphery individuals in the network. Our empirical analysis shows that core users exert more influence on periphery users than vice versa. Our findings provide important insights to help identify influential members in the social network, which is potentially useful to design optimal targeting strategies to improve current network-based marketing practices.*

# 1 Introduction

Mobile handset is the central medium through which consumers subscribe to mobile network operators (MNOs) traditionally for the purpose of using communication services such as calling and text messages. As the number of mobile subscriptions worldwide is passing 7 billion by the end of 2015 (ITU 2015), smartphones have appeared as the standard configuration for mobile handsets once dominated by feature phones with fixed functionalities at the time of manufacture. Gartner (2015) estimated that annual smartphone sales have totalled over 1 billion units and exceeded that of feature phones since 2013. On one hand, the development of smartphones has fuelled the rapid growth in creating multi-sided technological and commercial platforms that represent the accelerated convergence of mobile telephony, Internet services and personal computing (Campbell-Kelly *et al.* 2015). The ecosystem of these platforms consists of interdependent actors, including chipset makers, smartphone vendors, MNOs, and mobile OS and application developers, who together contribute complementary innovations that empower the "over-the-top" applications, such as web browsing, video streaming, online gaming, mobile banking, *etc*. These services are easily installed in the smartphone as mobile apps that consumers can choose based on own preferences and thus have a significant impact on consumer welfare (Ghose and Han 2014). Meanwhile, the smartphone market has witnessed new entrants such as Apple and Samsung outcompeted the incumbents for their superior capabilities to engage with a variety of stakeholders across the value chain (Pon *et al.* 2014). This resulted in the proliferation of smartphones both in terms of number of new models and high variations among heterogeneous manufacturers, implying that product differentiation still characterize this innovative and competitive market (Cecere *et al.* 2015). On the other hand, total service revenues in wireless industry nowadays remained steadily growing. This can be in particular attributed to the expanded smartphone penetration, in that the greater value associated with smartphone users, as opposed to ones equipped with feature phones, is reflected by their higher willingness to shift to more expensive tariff plans with mobile broadband access as add-on service (OECD 2013). However, MNOs experienced the revenue gap caused by the continued decline in voice and SMS usage and heavy investment in network resources to handle capacity issues, due to the explosion of mobile data traffic. McKinsey & Company (2014) reported that global mobile data traffic increased nearly 40 times from 2008 to 2013, whereas the generated revenue streams from it merely tripled. To ensure the profitability and sustainable growth, MNOs have to devise effective strategies to keep inducing subscribers to upgrade to newer generation of smartphones, and meanwhile transition their tariff structures to become more data-centric. More importantly, the popularity of smartphones stimulated the emergence of innovations that are transforming many industries for immense economic and social development. Therefore, as one of the most dynamic market segments in wireless industry, understanding adoption patterns of smartphones is of great interest to not only handset manufacturers and MNOs, but also to service providers and application developers.

Factors that influence consumers' acceptance and adoption behavior of mobile services have long been a fundamental research theme in mobile business community and mobile computing within the IS discipline at large, mostly dominated by conceptual approaches (Budu and Botateng 2013; Ladd *et al.* 2010). For example, Tscherning and Damsgaard (2008) designed a holistic framework to analyze scholarly literature about diffusion of various types of mobile communication products and services, and found that most existing research employed either Technology Acceptance Model (TAM) (*e.g.* Sarker and Wells 2003) or Diffusion of Innovation (DOI) theory (*e.g.* Kakihara 2014). López-Nicolás *et al.* (2008) linked these two theoretical frameworks together, showing that social factors are strongly related to consumer's perceived benefits to adopt advanced mobile services, because DOI theory complements the TAM in the way that explained the potential adopters' attitudes towards mobile innovations are affected by the information received from their social environment. However, in the case of smartphone adoption, empirical studies on how the focal subscriber decides to adopt the smartphone, because she is informed about it, through the exposure from friends who have already adopted in the social neighborhood, are still limited. This is mostly because large-scale mobile dataset that can be used to construct mobile social network (MSN) and establish the relationship between peer influence and smartphone adoption is lacking until recently (Blondel *et al.* 2015). Two exceptions include one study considering smartphone as a general product category (Risselada *et al.* 2014) and the other focusing on one specific model of Apple's iPhone 3G (Matos *et al.* 2014). Both studies confirmed that peer influence have a positive impact on consumer's decision to adopt the smartphone.

In this paper, we contribute to the existing literature mainly from two aspects: First, we leverage network structure and specific position of each individual in the social network to account for and measure heterogeneous role of peer influence in smartphone adoption. We do so by introducing core-periphery structure as a meso-level organizational principle of network to complement the use of centrality measures that are derived from either global properties of the network (macro-level) or individual's local social neighborhood (micro-level). Second, we unify overlapping community structure with core-periphery structure, by specifying positive correlations between the overlaps of the community affiliations and the coreness of individual's network position. Our empirical results on iPhone 3G adoption, given the presence of confounding factors, demonstrate the asymmetric peer influence between subscribers positioned at the core and at the peripheral parts of the network, i.e. core users may exert stronger influence to periphery users than vice versa. We believe our findings provide important insights on identifying influential members in the social network, which is potentially useful to design optimal targeting strategy to improve current network-based marketing practices.

## 2 Related Literature on Peer Influence and Product Adoption

Peer influence via interpersonal communications, either online or offline, has been acknowledged the important role in determining consumer's choices for decades. In general terms, when we speak of influence within a social network, we assume that controlling all other factors, there exists temporal dependence between the focal individual's behavior and her direct neighbors' previous behavior (e.g. Bass 1969; Roger 2003). For example, marketers increasingly realize to harness the power of social links between consumers to complement traditional advertising strategy (Bruce *et al.* 2012; Trusov *et al.* 2009), aiming at triggering viral adoptions through social contagion (Van der Lans *et al.* 2010).

Traditional diffusion studies (e.g. Bass model and its variations) typically use aggregated data to model product growth at the market level without detailed network information, so that peer influence is simplified as just to estimate one relatively parsimonious parameter (Hartmann *et al.* 2008). Given that modern IS diligently record digital traces of human communications, such as emails, phone calls, tweets, *etc.*, at an unprecedented scale, the resulting massive data of social interactions can facilitate researchers to investigate peer influence with new insights. Peres *et al.* (2010) critically reviewed the shift from aggregate-level modelling to individual-level product adoption from perspective of consumers, and they noted the increasing contributions from observing and measuring contagion effect on actual social networks.

However, empirical analyses on identifying peer influence using observational networked data still face significant challenges, as it is known that estimation is subject to several confounding factors, such as homophily, correlated unobservable and simultaneity, among others (Aral 2010; Manski 1993). Due to the endogenous formation of social network, the correlated behavior among connected individuals can be ascribed to both influence and their inherent similarities - homophily (McPherson 2001). Thus misattribution of homophily to influence may lead to significant overestimation of the latter (Aral *et al.* 2009; Davin *et al.* 2014; Tucker 2008). Another source of correlation stems from the unobserved external stimuli that drive actions of connected individuals similarly. One example of correlated unobservable in the case of product adoption is mass marketing and/or media exposure that may potentially bring the awareness and interests of the product to everyone in the same fashion. For example, Van den Bulte and Lilien (2001) re-analyzed the classic study on diffusion of medical innovation by (Coleman 1966) that articulated the strong influence in doctor's new drug adoption decision, but found no evidence of contagion effect after controlling for marketing efforts. Finally, the simultaneity issue indicates the contemporaneous interdependence between the focal individual and her peers. Manski (1993) referred to this as "reflection problem" and it may also lead to upward bias in the estimation (Nair *et al.* 2010).

Numerous scholars have contributed to the toolkits for the purpose of quantitatively estimating peer influence in social networks given the presence of confounding factors, including actor-based model families (Lewis *et al.* 2012; Steglich *et al.* 2010), fixed or random effects models (Bramoulle *et al.* 2009; Goldsmith-Pinkham and Imbens 2013), hierarchical Bayesian model (Ma *et al.* 2014), instrumental variable methods (Tucker 2008), latent space model (Davin *et al.* 2014), matched sample estimation (Aral *et al.* 2009; Han and Ferreira 2014), randomization test (Anagnostopoulos *et al.* 2008; La Fond and Neville 2010), to name just a few. Despite the sophistication of these model specifications on handling observed homophily, Shalizi and Thomas (2011) argued that peer influence and homophily are generically confounded observationally and can not be readily disentangled from each other, because latent homophily may still remain as a component of the estimated influence.

Moreover, even when researchers explicitly address the endogeneity issue through robust identification strategies that alleviate bias from confounding factors, the estimated effects of peer influence may fail to account for individual heterogeneity in the tendency to influence (or be influenced by) peers, i.e., some consumers exert disproportionate influence to others and vice versa. As such, the notions of influentials or opinion leaders that catalyze product diffusion have received considerable attention with both theoretical backgrounds (Katz and Lazarsfeld 1955; Roger 2003) and empirical evidences (Goldenberg *et al.* 2009; Iyengar *et al.* 2011). However, Watts and Dodds (2007) doubted the influential hypothesis and found that cascades of influence are largely driven by "a critical mass of easily influenced individuals". Van den Bulte and Joshi (2007) introduced a two-segment structure with asymmetric influence in the diffusion model: one segment of influentials who affect another segment of imitators whose adoption can hardly affect influentials, and they showed that this two-segment approach fits data better than the standard mixed-influence model. Likewise, such a structure is consistent with several empirical studies that discover asymmetric peer effect in adoption behavior between influential and susceptible individuals (Aral and Walker 2012; Iyengar *et al.* 2011; Nair *et al.* 2010; Tucker 2008). These findings shed light on the new research direction towards understanding which individuals and in what way should be targeted so as to achieve mass adoption through contagion (Aral 2010).

Canonical methods to identify opinion leaders typically fall into three categories: 1) individual's authority is predetermined in a formal organizational environment (*e.g.* managers *vs.* workers) (Tucker 2008), which can hardly be adapted to common scenarios; 2) individuals are surveyed to directly nominate opinion leaders in their reference group (Nair *et al.* 2010); 3) individual's status is based on certain socio-metric techniques, by calculating network centrality scores after capturing self-reported social interaction information (Banerjee *et al.* 2013). Iyengar *et al.* (2011) compared the relationship between self-reported and socio-metric leadership and found that not only these two types of opinion leadership are barely moderately correlated, but also adoption patterns are distinct from each other. Meanwhile, how network structure and individual's positions in the network moderate the contagion process remains elusive. Needless to say, none of these methods can be well suited to generalize into large-scale network settings.

The inclusion of full network structure provides new insights when assessing heterogeneous effects of peer influence. Jackson *et al.* (2015) broadly described a taxonomy of "macro"-level and "micro"-level characteristics emerged from social network that can affect product diffusion. On one hand, the macro-level patterns relate to the global properties of the network, such as network densities, degree distributions, path lengths, and so on. These summary statistics capture the essential graph topologies of the network and have seen as primary determinants of process of diffusion (e.g. Roger 2003; Jackson and Rogers 2007). However, such overall characterizations of network overlook the richness of dyadic social interaction information between pairs of individuals. On the other hand, the micro-level patterns refer to the local network structure among individual's connections. Sundararajan (2008) explained micro patterns arise because individuals know their own immediate social neighborhood well and are likely to know less about the structure of their neighbors' local network, but have minimal knowledge about the rest of the network. Hence individuals with large number of connections (high degree) and/or are strongly interconnected with each other (high clustering) are assumed to serve like a hub in diffusion process, because they are more likely to receive and pass the information (Goldenberg et al. 2009). It is just unsurprising that highly central individuals who occupy the key positions are deemed as potential influentials, so that various centrality measures have been used to represent different aspects of importance in a myriad of applications (Jackson 2014). However, when it comes to learning about how individual's position matters, relevant centrality measures may require information that extends beyond individual's own local network structure, i.e., individual's centrality is recursively related to the centralities of her neighborhoods. Therefore, Jackson et al. (2015) pointed out the "blur" between macro and micro measures of network structure and individual position that needs further exploration.

Lastly, the underlying distinctive generating mechanisms of correlated behavior resulted from confounding factors such as homophily and heterogeneous influence may lead marketers to implement different strategies in order to boost product sales. If homophily is the primary driving force, then targeted actions should be placed upon segmented individuals with similar characteristics (Hill et al. 2006), while if influence is in effect, then marketers should focus on influential individual (Iyengar et al. 2011), or engineer products with attributes that can augment viral adoptions (Aral and Walker 2011). If influence and homophily are intermingled, as in most cases where social network is formed as a result, then the optimal strategy should be to target influentials of each segmentation and design incentives that foster the adoption across segmentations (Aral 2010). Therefore, without appropriately accounting for the relative importance and interplay between peer influence and homophily as well as the individual heterogeneity being influential and susceptible to influence does not only complicate the identification issue, but also hamper the confidence of marketers to measure and manage the effect of peer influence more systematically in order to ensure desirable outcome from their marketing efforts.

## 3  Background and Data Description

We obtain the EURMO dataset from a major MNO, which includes call detail records (CDR) for over 5 million subscribers between August 2008 and June 2009 in one European country. Subscribers are identified by their anonymized phone numbers. For each call we know the initiator and the recipient, the timestamp, and the GPS coordinates of the connected cell tower. By aggregating GPS coordinates over the entire period, we can approximate subscriber's home location as where they spend most of their days at municipal level[1]. We further infer the socio-economic information (e.g. wage) by cross-referencing the latest census. We also have a set of subscriber characteristics such as gender and usage history since their subscription to EURMO, which includes tariff plan, handset usage and supplementary services (e.g. mobile Internet). Table 1 lists relevant extracted variables with types and brief descriptions.

| Variable | Type | Description |
| --- | --- | --- |
| *gender* | categorical | Self-reported gender (male, female, unknown) |
| *wage* | categorical | Inferred wage level (very low, low, average, high, very high) |
| *prepaid* | binary | Prepaid tariff plan (yes, no) |
| *phone_technology* | cateogrical | Handset technology (2G, 2.5G, 3G, 3.5G, other) |
| *mobile_internet* | binary | Mobile broadband as add-on service (yes, no) |
| *phone_age* | continuous | Age of currently owned handset (year) |
| *tenure* | continous | Tenure since subscription (year) |
| *region* | categorical | Home location at municipal level |

Table 1.     List of covariates extracted from EURMO.

We decide to choose Apple's iPhone 3G (referred only as iPhone hereafter for brevity's sake) as the exemplary model to examine the role of peer influence in its adoption for the following reasons. First, compared to its predecessor and other smartphones, iPhone includes several innovative features that may incur contagious adoption: 1) it supports tri-band UMTS/HSDPA (also dubbed 3.5G) and enhanced browsing functionality that allow faster and easier mobile internet access, which are likely to generate positive network effects (West and Mace 2010); 2) at the day before its release, Apple introduced the App Store from which users can download 500 third-party applications and over 15,000 after the first 6 months and many of these apps are social in nature (e.g. those released by social networking sites) (Davin *et al.* 2014); 3) beyond technical utility, iPhone may also provide its users with hedonic benefits, e.g. the haptic experience through the touch-screen technology, which may be appealing to consumers with social needs (Arruda-Filho et al. 2010). Second, EURMO is the sole partner with Apple with exclusive arrangement. Meanwhile, CDR data spans from the month right after the release of iPhone and to the month before that of the successive model iPhone 3GS, so that we are able to capture the full cycle of the adoption. Third, iPhone is perceived as costly goods in this market. For potential adopters, opinions about this product from already adopted friends are likely to be as important, if not more, as when adopting cheap mobile product categories (e.g. caller ringback tone (Ma *et al.* 2014) or mobile apps (Davin *et al.* 2014)), in order to mitigate their uncertainty and justify the cost associated with the adoption. In our period of analysis, there are 20,570 iPhone adopters with complete profiles.

---

[1] The municipal location is defined as Nomenclature of Units for Territorial Statistics (NUTS) III, which is a geocode standard across European countries by Eurostat for statistical purposes.

We use CDR to construct the MSN as an undirected call graph. Specifically, we denote two subscribers to befriend each other if they exchange at least one call in the same calendar month. The mutual relationship between subscribers ensures that we preclude communications that are unlikely to represent the social ties, such as customer services and PBX machines. The resulting network consists of 5,535,388 subscribers and 66,717,468 edges with mean, standard deviation, and median of degree being 24.1, 25.7 and 16, respectively. Figure 1 shows that the degree distribution of MSN is highly skewed and heavy tailed. This implies that individuals have heterogeneous degree of social ties, such that they may have different roles in moderating product adoption through the peer influence.

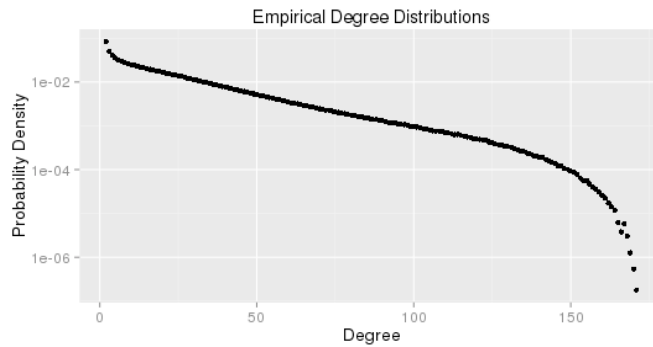

*Figure 1.    Empirical degree distributions of MSN generated from EURMO's CDR.*

## 4    Discovering Social Circles in Ego Network

Recent empirical analyses on statistical properties of real world MSN provide evidence of community structure embedded within the network (Girvan and Newman 2002; Park and Barabási 2007). Individuals tend to form social ties and become closely connected because of their homophilous characteristics. Such selection process produces structural consequences for the network that mainly consist of cohesive groups of similar individuals. Thus it is natural to bind community-aware approach with the estimation techniques of peer influence, i.e., to establish the existence of these groups and to assign group affinities to each individual into the chosen model, in order to control for the group-level unobserved heterogeneity (Shalizi and Thomas 2011).

The community structure in MSN does not only validate the theoretic role of homophily and influence in tie formation, but also provides several important insights into the task of community structure inference, when we only observe the resulting networks as follows. First, uncovered communities should exhibit real social meaning, as individuals in the same community have some natural affinity for each other or some fundamental characteristics in common. Meanwhile, they should be more likely to connect to each other than those who belong to different communities. Hence community discovery method should entail two different sources of information together, i.e., individual characteristics and social connections among them. Second, in many actual networks, individuals may belong to multiple overlapping social circles (Ahn *et al.* 2010; Palla *et al.* 2005), such as families, college friends and co-workers, etc. This is also aligned with the notion of pluralistic homophily across different social dimensions. Third, as also noted in Shalizi and Thomas (2011), misspecification of community structure (e.g. simple modular and/or disjoint structure) may even worsen the problem and lead to biased model estimation. Fourth, as the complexity of network structure grows exponentially with the size, computational costs still remain challenging for large-scale networks. Therefore, instead of exploring the whole network, extracting subpopulation through community discovery does not only significantly reduce group-level heterogeneity that may potentially confound the result, but also help lessen the computational cost (Zhang *et al.* 2011).

**Subpopulation Extraction** We resort to state-of-the-art method of discovering Communities from Edge Structure and Node Attributes (CESNA) to extract the subpopulation from the MSN[2]. CESNA is a community discovery algorithm that considers individual characteristics, network structure as well as interactions between these two sources of information. It can detect overlapping communities with high accuracy and scalability over many existing community detection algorithms, particularly on large-scale networks. For sake of space, details beyond the mechanics of CESNA can be found in (Yang *et al.* 2013), and we only note the following implementation procedures[3]: i) for each iPhone adopter, we construct the ego-network that contains adopter and their direct neighbors; ii) for each subscriber in the ego-network, we extract a list of 0-1 valued covariates specified in table 1 that represent pluralistic homophily, including gender and wage (socio-demographic homophily); tariff plan, phone technology and mobile broadband (contextual homophily); and home location (spatial homophily); iii) we apply CESNA on each ego-network using both node and edge information with the optimal number of communities identified through cross-validation; iv) we remove duplicated and nested communities and only retain those having iPhone adopters. As a result, we obtain 11,454 communities with 202,743 subscribers, 14,685 (71%) of which are iPhone adopters, and nearly half of the original network size is dropped.

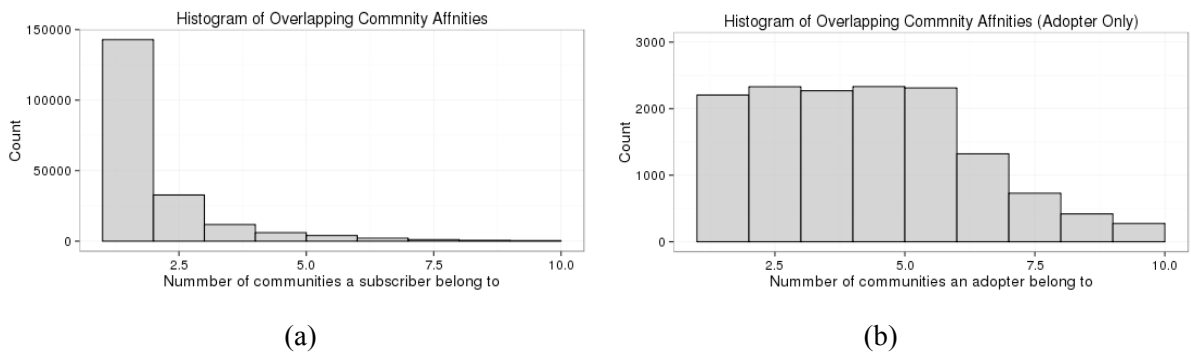

*Figure 2.    Histogram of overlapping community affinities for all subscribers (left) and iPhone adopters (right).*

**Core-Periphery Structure** As Fig. 2a shows, from the extracted subpopulation, over 70% of subscribers belong to only one community and nearly 90% of those belong to two, while only about 5% of subscribers belong to more than 5 communities. However, we find clearly different patterns of community memberships for iPhone adopters alone (Fig. 2b), that they are more likely to belong to multiple communities. This provides us with extra implications that iPhone adopters tend to link with others with the shared properties through overlapping social circles. Moreover, we believe that the intersection of overlapping communities may reveal another type of "meso"-level organizing principle of network: *core-periphery structure* (Yang and Leskovec 2014). In general, *core nodes* refer to set of central nodes that are connected to other core nodes as well as peripheral nodes, while *peripheral nodes*, by contrast, are only loosely connected to the core nodes but not to each other (Borgatti and Everett 2000). In this regard, following the measure proposed by Yang and Leskovec (2014), we validate the existence of a global core-periphery structure in our subpopulation (see Fig. 3).

---

[2] We do not aim to survey the rich literature on community discovery methods as it is beyond our objective (refer to e.g. Fortunato 2010) for comprehensive reviews.
[3] The source code is available at http://snap.stanford.edu. We also discussed with authors of CESNA about implementation details through private communications.

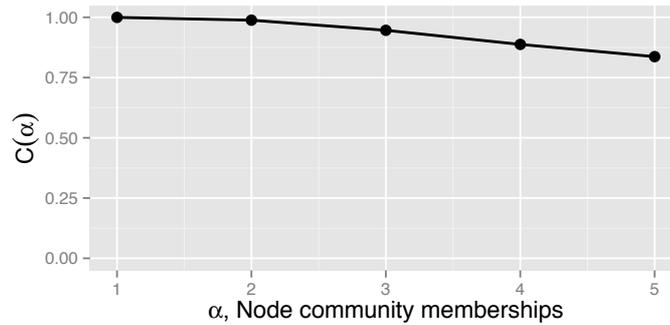

*Figure 3.    The fraction of nodes $C(\alpha)$ in the largest connected component of the induced subgraph on the nodes that belong to at least one community. A high $C(\alpha)$ means that there is a single dominant core.*

Core-periphery structure captures individuals' network positions that current centrality measures do not account for. For example, if individuals with high degrees (hubs) exist at the periphery of a network distant from densely connected core, they will still have insignificant impact in the spreading process as adoptions are likely to be confined to their affiliated communities, whereas less connected individuals who are strategically positioned at the overlaps of communities and thus become more central, then adoptions may percolate across communities and to other central individuals who are also placed at the overlaps of communities and so on (Aral 2010). Therefore, we argue that individuals who are placed at the core of the network are more likely to be influential compared to those at rather isolated peripheral parts, such that peer influence between core and periphery members of the network may appear to be asymmetric.

## 5  Empirical Results

We describe our empirical approach to estimate the peer influence on iPhone adoption as follows. We organize the subpopulation data into a panel where each individual is a subscriber and each period is a calendar month and observations after the first adoption are removed from the sample. In this way, we can estimate the discrete-time hazard model using standard binary choice specification (Allison 1982). The dependent variable is an indicator for when a subscriber first starts to use iPhone. Our model specification includes subscriber-specific characteristics such as social-demographic indicators, wireless technological aptitude, service usage and cumulative adoptions from friends that are deemed as either core or periphery. We denote a subscriber as core if she belongs to at least 5 communities and as periphery if otherwise. Among 9,194 core nodes, 5,548 (60%) are iPhone adopters, whereas only 5% of peripheral nodes are adopters. With panel data, we are also concerned about controlling for the unobserved heterogeneity as much as we can, so we introduce dummy variables to control for fixed effects across home region, time and community memberships. These variables can help reduce systematic differences across locations where subscribers may have different experience of smartphones for using mobile broadband because urban areas normally support better network coverage, across time of periods due to the seasonal effects (e.g., campaigns during Christmas), and across communities for common traits at group level that we explained in the preceding section. Table 2 summarizes the descriptive statistics for covariates used in regressions.

| Variable | Description | Mean | Std |
|---|---|---|---|
| Dependent variables | | | |
| Adopted_core$_t$ | Indicator variable for first month a core subscriber adopts the iPhone | 0.086 | 0.28 |
| Adopted_peri$_t$ | Indicator variable for first month a peripheral subscriber adopts the iPhone | 0.004 | 0.065 |
| RHS variables | | | |
| Gender_male | Indicator variable for male subscriber | 0.228 | 0.419 |
| Gender_female | Indicator variable for female subscriber | 0.164 | 0.37 |
| Gender_other | Indicator variable for subscriber who did not report gender | 0.608 | 0.488 |
| Prepaid | Indicator variable for subscriber who used prepaid tariff plan | 0.471 | 0.499 |
| Phone_2G | Indicator variable for subscriber who used 2G handset | 0.122 | 0.327 |
| Phone_2.5G | Indicator variable for subscriber who used 2.5G handset | 0.479 | 0.499 |
| Phone_3G | Indicator variable for subscriber who used 3G handset | 0.355 | 0.479 |
| Phone_3.5G | Indicator variable for subscriber who used 3.5G handset | 0.039 | 0.193 |
| Phone_other | Indicator variable for subscriber who used an unknown ranged handset | 0.005 | 0.07 |
| Mobile_internet | Indicator variable for subscriber who used mobile broadband | 0.036 | 0.187 |
| Phone_age | Number of months subscriber have used currently owned handset | 0.915 | 0.695 |
| Core_frd | Number of subscriber's friends who deemed as core members | 1.70 | 2.57 |
| Peri_frd | Number of subscriber's friends who deemed as peripheral members | 14.46 | 10.18 |
| Core_frd_adopt$_{t-1}$ | Cumulative adoption by subscriber's friends deemed as core members | 0.609 | 0.919 |
| Peri_frd_adopt$_{t-1}$ | Cumulative adoption by subscriber's friends deemed as peripheral members | 0.508 | 0.692 |
| Tenure$_t$ | Number of months since the subscription to EURMO | 6.61 | 4.43 |
| Control for regions | Dummies for each NUTS III region | | |
| Control for month | Dummies for each calendar month from August 2008 to June 2009 | | |
| Control for community | Dummies for each community membership | | |
| Total observations | | 2,116,855 | |

*Table 2.     Summary statistics of covariates used in regressions. Time-invariant covariates are measured on June 30, 2008, the last day of the month before the release of iPhone.*

**Identification Strategy** Still, unobserved heterogeneity such as latent homophily might bias our estimation. Matos *et al.* (2014) described the issue as "one-hop homophily", because it is assumed to plays a role in correlated outcome only between direct contacts but not over two hops or more. So they argued that for a pair of connected subscribers, one has a third friend who is not a friend with the other, then this third friend's decision to adopt iPhone is correlated to the pair only through the one she is connected with. Bramoulle *et al.* (2009) illustrated this type of network structure as "intransitive triads" and they showed that instrumental variables (IV) built from it can sufficiently identify peer influence. Therefore, we follow the same IV approach to alleviate the endogeneity concerns.

Before we delve into the 2-stage IV approach, we first present results from pooled Probit model where we stratify subscribers based on core/peripheral network positions and separately measure how they responds to adoption by their core and peripheral friends. This allows a naive exploration of whether the network position matters for individual's heterogeneity to exert and receive peer influence. The result for the simple model is shown in column [1] and [3] of Table 3. The likelihood of subscriber to adopt iPhone is positively associated with the cumulative friends' adoption. More importantly, we clearly see *asymmetric* peer influence between the focal subscribers and their core/periphery friends. Specifically, all subscribers are more likely to get influenced by their core friends, regardless of their own network positions. More interestingly, periphery subscribers are more likely to get influenced by core friends than vice versa. This suggests that subscribers who occupy the central positions are likely to be more influential, while those who are located at the peripheral parts of the network are more susceptible to influence from the core friends.

Next, we estimate the model using 2-stage residual inclusion (2SRI) to deal with endogeneity issues. Compared to 2-stage predicator substitution (2SPS), which is widely used in linear regression models, 2SRI is similar in the first stage, whereas endogenous regressors and the first-stage residuals are included in the second stage. In nonlinear setting, 2SRI may yield consistent results whereas 2SPS is not (Terza *et al.* 2008). As explained earlier, we use "cumulative adoptions by friend of friend not friend of the focal subscriber" as the IV. The results obtained from 2SRI model is listed at the column [2] and [4] of Table 3[4]. Our findings are largely consistent compared to the simple pooled Probit model, while asymmetric peer influence between core and periphery subscribers are even more significant. This further renders us more confident about how network structure and positions would impact the focal subscribers' behavioral changes from the heterogeneous influence received from their friends.

|  | Core | | | | Periphery | | | |
| --- | --- | --- | --- | --- | --- | --- | --- | --- |
|  | Probit [1] | | 2SRI [2] | | Probit [3] | | 2SRI [4] | |
| $Core\_frd\_adopt_{t-1}$ | 0.125*** | (0.007) | 0.318*** | (0.04) | 0.200*** | (0.013) | 0.706*** | (0.032) |
| $Peri\_frd\_adopt_{t-1}$ | 0.042** | (0.012) | 0.074*** | (0.015) | 0.047*** | (0.013) | 0.141*** | (0.009) |
| $Core\_frd$ | 0.03*** | (0.002) | 0.036*** | (0.002) | 0.099*** | (0.003) | 0.109*** | (0.004) |
| $Peri\_frd$ | 0.005*** | (0.0006) | 0.013*** | (0.003) | 0.016*** | (0.0004) | 0.025*** | (0.001) |
| $Gender\_male$ | 0.116*** | (0.022) | 0.104*** | (0.022) | 0.153*** | (0.009) | 0.163*** | (0.01) |
| $Gender\_female$ | -0.024 | (0.03) | -0.018 | (0.031) | -0.055*** | (0.013) | -0.053*** | (0.013) |
| $Prepaid$ | -0.129*** | (0.022) | -0.035* | (0.02) | -0.054*** | (0.009) | -0.016* | (0.01) |
| $Phone\_2.5G$ | 0.317*** | (0.046) | 0.316*** | (0.046) | 0.317*** | (0.018) | 0.331*** | (0.018) |
| $Phone\_3G$ | 0.378*** | (0.046) | 0.338*** | (0.047) | 0.435*** | (0.018) | 0.452*** | (0.018) |
| $Phone\_3.5G$ | 0.461*** | (0.050) | 0.439*** | (0.051) | 0.679*** | (0.022) | 0.718*** | (0.022) |
| $Phone\_other$ | 0.415*** | (0.096) | 0.501*** | (0.099) | 0.504*** | (0.049) | 0.547*** | (0.049) |
| $Mobile\_internet$ | 0.132*** | (0.023) | 0.125*** | (0.023) | 0.125*** | (0.017) | 0.142*** | (0.018) |
| $Phone\_age$ | -0.029* | (0.012) | -0.031* | (0.01) | -0.048*** | (0.006) | -0.045*** | (0.006) |
| $Tenure_t$ | 0.08*** | (0.016) | 0.08*** | (0.02) | -0.05*** | (0.01) | -0.04*** | (0.01) |
| Control for regions | Yes | | Yes | | Yes | | Yes | |
| Control for month | Yes | | Yes | | Yes | | Yes | |
| Control for community | Yes | | Yes | | Yes | | Yes | |
| Pseudo $R^2$ | 0.0751 | | 0.0889 | | 0.0657 | | 0.0869 | |
| Observations | 63,863 | | 63,863 | | 2,052,992 | | 2,052,992 | |

***p<0.001, **p<0.01, *p<0.05 (all tests are two tailed)

*Table 3.    Regression results obtained from Probit and 2SRI models*

# 6  Discussion

This paper adds to the growing literature that explores the role of peer influence in product adoption using individual-level observational data with detailed large-scale network information. We link two fundamental organizational principles of social networks: community and core-periphery structure, by showing that an individual in the core is also more likely to show up in more overlapping communities. Furthermore, our results provide evidence of the asymmetric effect of peer influence between core and periphery users. In particular, core users exert more influence over periphery users than vice-versa. Our findings suggest that taking both homophily and influence into account, targeting subscribers who belong to many communities and thus are more likely to be located at the core of network may help spread the product within communities, as users in the same community tend to exhibit similar traits, as well as across communities through closely connected core subscribers who are likely to receive multiple referrals and thus generate a "social multiplier" effect (Hartmann *et al.* 2008). In our future work, we will perform simulated seeding strategies to have a better understanding of the benefits associated with our findings.

**Acknowledgement**  This work was partially supported by Portuguese Foundation for Science and Technology through Carnegie Mellon Portugal Program under Grant SFRH/BD/51153/2010.

---

[4] First-stage regression results are not reported here due to the space limit but are available upon request.